# Is Female Health Cyclical? Evolutionary Perspectives on Menstruation

Alexandra Alvergne[1] and Vedrana Högqvist Tabor[2,3]

1-School of Anthropology, University of Oxford, UK

2-Clue by BioWink, Berlin, Germany

3-Boost Thyroid by VLM Health, Berlin, Germany

# Summary

**Why do some females menstruate at all? Answering this question has implications for understanding the tight links between reproductive function and organismal immunity. Here we build on the growing evidence that menstruation is the by-product of a "choosy uterus" to (i) make the theoretical case for the idea that female immunity is cyclical in menstruating species, (ii) evaluate the evidence for the menstrual modulation of immunity and health in humans and (iii) speculate on the implications of cyclical female health for female behaviour, male immunity and host-pathogen interactions. We argue that an understanding of females' evolved reproductive system is foundational for both tackling the future challenges of the global women's health agenda and predicting eco-evolutionary dynamics in cyclically reproducing species.**

**Highlights:**

- Accumulating evidence suggests that the menstrual cycle is underpinned by inflammatory patterns that change in a way that enables maternal control over the selection of viable embryos.

- In humans, albeit results are mixed due to methodological issues, naturally cycling females exhibit menstrual variation in immune function and non-reproductive health.

- Female cyclical immunity ought to be considered for understanding phenotypic diversity in female behavioural immunity and reproductive behaviour, male immunity and the evolution of sex-specific pathogen virulence.

**Word count**: 3478

# The Evolutionary Origin of Cyclic Decidualization

Why do some female mammals menstruate at all? Growing evidence from human data suggests that the menstrual cycle is underpinned by inflammatory patterns that change in a way that enables maternal control over the selection of viable embryos [1,2]. If immunity does indeed change over the menstrual cycle, what are the implications for the fitness of individuals and for population processes? In this paper, we show that understanding how the menstrual cycle regulates the immune system has potentially far-reaching implications not only for improving our appreciation of the determinants of health beyond reproductive health but also for predicting ecological dynamics and pathogen evolution.

It is widely acknowledged that menstruation is rare among placental mammals, although it depends on how menstruation is defined [3]. In dogs, bleeding correlates with ovulation, originates in the vagina (not the uterus) and in the absence of a pregnancy, the endometrium is reabsorbed. In humans, by contrast, menstruation results from the cyclical shedding of the inner lining of the uterus and occurs when a successful pregnancy has not been established. This human pattern, profuse uterine bleeding that occurs in the absence of pregnancy and after ovulation, is only observed in a few mammal species including humans, some old world primates and a few species of bats and rodents (Box 1, [3–5]). In addition, menstruation is potentially nutritionally and reproductively costly [3], although this view has been challenged [6,7]. Thus, why has menstruation evolved at all has puzzled evolutionary biologists for more than 20 years (Box 1).

There is now growing evidence that menstruation is the by-product of uterine evolution in response to foetal-maternal genetic conflict over maternal investment [4], a particularly intense phenomenon in humans due to both trophoblast invasiveness and a high rate of genetic abnormalities [2,4]. In humans and other menstruating species, the "preparation" of the endometrium for pregnancy (i.e. the decidualization process), is under maternal rather than embryonic control: it occurs every cycle in the absence of embryonic cues (i.e. it is spontaneous). The process of spontaneous decidualization is achieved through a bi-phasic immune response during the luteal phase [2,8]: first, an acute inflammatory response to select embryos with a potential of surviving and second, a profound anti-inflammatory response to enable implantation (Figure 1). In non-conceptive cycles, the response is triphasic as it further includes menstruation, a massive inflammatory event triggered by falling progesterone levels, which, through the recruitment of stem cells, enables the regeneration of the endometrium [8].

The evolution of spontaneous decidualization coupled with cyclic menstruation has been hypothesized to be a major evolutionary adaptation that optimizes the balance between embryo selectivity and embryo receptivity [1,2,4,9]. A well-balanced decidual cell reaction (DCR) is needed to maximise reproductive success [2]: a bias towards embryo selection (an excessive DCR) reduces the window of opportunity and likelihood of pregnancy while a bias towards embryo implantation (an impaired DCR characterized by high levels of pro-implantation cytokines) leads to both high fecundity and recurrent pregnancy loss [10]. A functional DCR is thought to benefit maternal reproductive outcomes by avoiding maternal investment in compromised embryos [2,9,10], but cyclical shifts in patterns of inflammation might also bring about costs for maternal non-reproductive health. Given that the immune system must trade-off resources between the fitness functions of immunity and reproduction during the luteal phase, is female health best conceptualized as cyclical in menstruating species?

One might argue that menstruation is too rare (i.e. in non-humans) or too evolutionary novel (i.e. in humans) for the concept of cyclical female health to apply to most menstruating species. Indeed, most female mammals spend a significant part of their reproductive lives pregnant or lactating. However, in non-human primates, females are known to experience periods of cycling referred to as "waiting time to conception" (e.g. from 4.7 to 26.9 months in wild Chimpanzees [11]). During that time, each ovulatory cycle is characterized by a luteal phase, which differs from non-menstruating species that typically require mating to induce ovulation and/or the differentiation of the endometrium [12]. The importance of menstrual cycling in natural fertility populations is also illustrated by the ubiquity of menstrual taboos and rituals in traditional high-fertility human populations [13]. Further, in westernized environments, there is now considerable variation in life history trajectories globally, with millions of women experiencing no or few childbearing events [14]. This shift in childbearing patterns correlates with a change in menstrual cyclicity: women experience either an increased number of menstrual cycles during their lifespan and/or altered menstrual cycles, whereby ovulation is suppressed following the use of hormonal contraceptives [15]. Understanding the evolutionary ecology of menstrual cycling is thus relevant across species and populations.

In this paper we ask whether female non-reproductive health is best conceptualized as cyclical in menstruating species. We review current knowledge on the menstrual regulation on systemic immunity, disease susceptibility and severity. We then speculate on the impact of menstrual cycling in immune function for female social behaviour, male immunity and host-pathogen interactions. We conclude by pointing to the need for studies on the ecology of menstruation to advance our understanding of the ecological determinants and evolutionary consequences of variation in cyclical immunity.

# The Evidence for Cyclical Health

This section reviews the evidence for cyclical female health focusing on the menstrual modulation of (i) inflammatory patterns and (ii) disease susceptibility, development and severity with regards to infections and chronic diseases. Note that data on the menstrual regulation of systemic immunity and non-reproductive health are scant and limited to humans (reviewed in [16,17]). While studies show that women's health is cyclical, methodological issues and the lack of recent meta-analyses prevent drawing a conclusive picture of phase-specific effects.

### *The cyclical modulation of systemic immunity*

The menstrual cycle is characterized by a shifting reproduction-immunity trade-off. At the proximate level, this trade-off is partly regulated through the actions of oestrogen and progesterone [18], two sex hormones produced by the ovaries and for which there are receptors on many immune cells [19]. It is generally accepted that estrogen has both inflammatory and anti-inflammatory properties, depending on the dose and the ratio between oestrogen and progesterone [20], and that progesterone is mainly anti-inflammatory [21]: a drop in progesterone levels acts as a trigger for the inflammatory event of menstruation [18]. At the ultimate level, a few studies have researched the hypothesis that the post-ovulatory phase of the menstrual cycle is characterized by a biased investment in reproduction at the expense of immune-competence.

A number of studies investigated the menstrual modulation of circulating concentrations of C-reactive protein (CRP), a systemic marker of inflammation used to quantify investment in immune-competence [22]. It yielded mixed results (Table 1). The phase most clearly associated with high CRP levels across studies is menstruation, which supports the view that menstruation is an acute inflammatory event [8]. However, CRP levels measured in the late luteal phase were found to be either higher, lower, or not significantly different from other phases. The association between levels of steroid hormones and CRP levels is complex [23] and opposite results have been reported (Table 1).

Other studies focused on the menstrual variation of the so-called "Th1/Th2" route, i.e. the relative importance of the immune response to intra-cellular pathogens (Th1 response, cell-mediated, deals with bacteria and viruses) and extra-cellular pathogens (Th2 response, humoral/blood, deals with self-tolerance and helminths) [24]. The production of Th2 anti-inflammatory cytokines assist in the maintenance of pregnancy and a defective Th2 response is linked to spontaneous abortions [25]. But knowledge of the menstrual regulation Th1/Th2 route is limited and based on small sample sizes. A study involving 13 women reported that a Th2 anti-inflammatory response was promoted in the luteal phase [26], a result also observed in an exploratory study of 14 sexually active American women [27]. Conversely, no evidence for menstrual variation of Th1 and Th2 mediators was found in a sample of American women [16] and no changes in Th1/Th2 ratio were observed in a sample of 22 healthy Korean women [28].

The results are inconsistent, and it is difficult to draw a firm conclusion. First, there is significant variation in steroid hormones across cycles [29,30], limiting the significance of results from studies drawing on one cycle only. Further, absolute steroid levels might not be informative [22]; rather, whether levels of estrogen and progesterone are rising or declining might be better cues for capturing changes in immunity. In addition, other variables such as ovulation status and sexual partnership, which predict whether or not CRP levels [22,31], the Th1/Th2 ratio [27] or other cytokines [31] cycle, should be considered. Fourth, studies sometimes lump together measurements taken in both the inflammatory and anti-inflammatory days of the luteal phase, preventing the detection of cyclical changes within the luteal phase.

### *The cyclical modulation of non-reproductive health*

While menstrual cycle research focuses on reproductive health, there is some evidence that the menstrual cycle influences non-reproductive health too. First, patterns of health and diseases vary between users and non-users of hormonal contraceptives (Box 2). Second, the menstrual cycle phase influences susceptibility to bacterial [32], viral [33] and fungal infections [19]. Infection susceptibility is heightened in the non-inflammatory phase of the cycle associated with high or rising levels of progesterone [19] (Figure 2b). For instance, Wira et Fahey [34] argue that there is a window of vulnerability for HIV infection in the 7-10 days following ovulation. However, while adult females develop higher antibody responses to vaccines than males [35], how the menstrual cycle influences the response to vaccination yet to be documented. Similarly, how the microbiome diversity changes during the menstrual cycle has received little attention [36].

Third, clinical data suggest that the menstrual cycle phase exerts a modulatory effect on some chronic diseases [17], such as inflammatory bowel diseases [37], asthma [38], epilepsy [39] and autoimmune connective tissues disorders [40]. During or next to the inflammatory phases of the menstrual cycle (ovulation and menstruation), symptoms are aggravated, while during the non-inflammatory phase (mid-luteal), symptoms are improved (Figure 2a). Due to the lack of recent data and meta-analyses, the picture is less clear for diabetes [41], cardiovascular diseases [42] or pain [43]. In addition, the cyclical experience of chronic diseases might be due to changing perception of disease severity, epigenetic regulation, the microbiome [36] and behaviour. Overall, there is little recent research on the menstrual regulation of chronic diseases despite its potential important implications for treatment. For instance, given the mitogenic effects of oestrogen and the known influence of steroid hormone receptors in the development and progression of numerous cancers [44], one might investigate whether cancer progression and treatment outcomes are modified by the menstrual cycle phase.

Finally, non-reproductive health might also be modulated by the cycling life-history (Figure 2). Beyond the known importance of reproductive scheduling for gynaecological cancers [reviewed in ,15], age at menarche correlates with the risks of both cardiovascular diseases [45,46] and some non-reproductive cancers (some non-Hodgkin lymphoma [47], lung [48], thyroid [49] and gallbladder cancers [50]; but not pancreatic [51], liver [52] or colorectal cancer [53]). It is unknown whether the link between the cycling life-history and health is mediated by lifetime exposure to hormones and/or inter-individual differences in hormonal levels. For instance, early age at menarche correlates with both increased exposure to and higher levels of steroid hormones [54]. Either way, accounting for lifetime exposure to cycles of inflammation or inflamm-aging [55] might shed new light on the importance of cycling immunity for non-reproductive health.

# Cyclical Female Health in Ecological and Evolutionary Contexts

Reproduction-immunity trade-offs in menstruating females are subject to both physiological and functional conflicts [18]. In addition to the physiological trade-offs imposed by seasonality in resource availability [56], in the second phase of the cycle, the immune system mu cope with the dual function of enabling the implantation of a healthy embryo while maintaining functionality in both selecting against compromised embryos [2] and defending against pathogens [18]. What are the ecological and evolutionary consequences of female cyclical investment in reproduction and immunity? In this section, we speculate on the salience of cyclical female health for understanding phenotypic diversity in female social behaviour, male immune function and the evolution of pathogen virulence.

*Behavioural immunity*

There is some indirect evidence that cyclical immunity activates the behavioural immune system, i.e. modulates social behaviour to compensate for variation in immune function. First, studies have shown that high levels of progesterone and/or the luteal phase correlate with decreased sexual activity in non-human primates [57] and in humans [58]. Second, research suggests that the menstrual cycle modulates prosocial behaviour. For instance, an online study showed that women were more cooperative in the early follicular as compared with the mid-luteal phase [59], and two recent studies found that progesterone levels correlate with disgust sensitivity [60,61] (but see [62]). However, absolute levels of steroids are not always good markers of the immune function, which might be otherwise more closely related to ovulation and sexual partnership [22]. Thus, studies on female behavioural immunity, otherwise hypothesized to account for diversity in the use of social information, appearance-based prejudices, xenophobia, sexual attitudes and dispositions towards gregariousness [63], should consider measuring not only sex hormones but also immune biomarkers when testing whether females avoid pathogen exposure when their immunity is compromised.

*Reproductive behaviour*

Little is known about the relationship between immunity and reproductive behaviour across the menstrual cycle, but existing data suggest that the relationship is bidirectional. First, sexual activity has been found to be reduced in the luteal phase in both old world primates [57] and humans [58]. However, whether the menstrual modulation of sexual behaviour results from cycling immunity because females are less proceptive when they are immunocompromised (behavioural immunity hypothesis) and/or from a shift in sexual attractiveness (ovulation detection hypothesis) is unknown. Second, Lorenz et *al.* showed that sexual partnership determines cyclicity in immunity: a shift is only observed among sexually active [22,27,31] women, which suggests that the immune system is plastic to reproductive opportunities. Those findings also raise the possibility that individuals who exhibit menstrual cyclicity in immune function, a potential cue for reproductive fitness if it correlates with the efficiency of the DCR, are more attractive and thus more likely to be sexually active. Further studies investigating the bi-directional relationship between immune cyclicity and reproductive behaviour within and between females present an exciting future prospect.

*Male immunity*

Life-history theory posits that individuals face a resource allocation trade-off between somatic and reproductive efforts [64]. Thus, if female immunity influences female sexual behaviour, then one might expect plastic adjustments in male immunity. In monogamous systems, if pairs engage in cyclical sexual activity then male somatic effort might be cyclically downregulated as a response to cyclical increases in male sexual activity. In polygynous mating systems, a rising number of naturally cycling females will shift the Operational Sex Ratio (i.e. OSR or the ratio of fertilizable females to sexually active males [65]) towards females, thus reducing male-male competition. On the one hand, the average male investment in immune function might increase if less energy is devoted to reproductive competition. On the other hand, increases in sexual activity afforded by a rising number of fertilizable females might rather drive male immunity down. The study of the interaction between female and male immunity might offer some insight into the flexibility of the immune system and the relevant time scale for its plastic adjustments.

*Pathogen virulence*

For parasites, natural selection operates on the rate of transmission between hosts [66], thus there is some scope for cyclical immunity to shape the evolution of pathogens: as compared to pregnant females, naturally cycling females might favour parasite transmission due to increased sexual activity and cooperative behaviour in the first phase of the cycle [59] while experiencing higher resistance to infection due to heightened immunity. If that is the case, one might expect cyclical immunity to shape the evolution of pathogen virulence (i.e. the reduction to host fitness due to pathogen growth within host) in several ways.

First, according to the virulence-transmission trade-off hypothesis [66], when opportunities for pathogen transmission increase, the fitness benefit of maintaining hosts alive is reduced and thus increased virulence is expected to be selected for [66]. Second, evolutionary ecological models have shown that interventions that lead to heightened immunity in some individuals (e.g. vaccinated individuals, naturally cycling individuals) might lead to the evolution of higher virulence in pathogens due to the increased resistance of hosts, and this might have dramatic effects on non-immune individuals (e.g. not vaccinated individuals [67], pregnant females). Finally, sex-specific virulence might evolve not because of variation in host immunity between males and females but because of differences in transmission routes between the sexes [68]. Female mammals are able to transmit pathogens not only horizontally but also vertically through pregnancy, birth and breastfeeding. If one assumes that the trade-off between virulence and transmission applies to horizontal but not to vertical transmission, virulence evolves to lower levels in females [68]. However, the situation might differ if either the rate of horizontal transmission increases in naturally cycling females or the rate of vertical transmission decreases in low fertility populations. Accounting for the cycling life-history might have implications for understanding differences in male-female fatality cases across populations, and the extent to which treatment should be sex-specific.

## Concluding remarks

Accumulating evidence suggests that, at least in humans, the female immune response is modulated by the hormones governing the menstrual cycle in a way that enables the implantation of a healthy embryo, even in the absence of fertilization. Although non-conceptive cycles are infrequent in non-human menstruating species, the occurrence of menstruation is on the rise in many contemporary human populations. Given the modulation of inflammatory patterns imposed by the menstrual cycle on the body, we argue that female health is best conceptualized as cyclical because symptoms and susceptibility to infection are expressed differently across the phases of the cycle, and women with different cycling life-histories vary in their susceptibility to and experience of diseases. Further, we contend that acknowledging the cyclical nature of female immunity in menstruating species might shed new light on behaviour, sex-specific immunity and the evolution of pathogen virulence. We now discuss the limitations of our approach and whether the concept of cyclical female health can be extended to cyclically reproducing species.

A significant limitation to our approach is the difficulty of studying the ecology of menstruation for understanding the causes of variation in patterns of menstrual cycling. Hormonal patterns are highly variable among populations, individuals and cycles (Box 3).

Thus, the study of the menstrual cycle requires both cross-sectional and longitudinal data. Recent years have seen a flourishing of period tracker apps for smart phones, and millions of women are now using them. Mobile phone apps offer a unique potential to document previously unknown phenotypic diversity. However, they are not yet ready for evolutionary ecological research. The pool of users is currently biased towards well-off WEIRD [69] people; they do not currently permit the tracking of many ecologically relevant variables (e.g. perception of social standing, stress, medical history, diet) although some are already recorded (e.g. sleeping and physical activity patterns); the complete cycling life-history (e.g. age at menarche, menstrual cycles, pregnancies, follow-up of contraceptive use) is rarely available and the in-built emotion categories are constrained by the biomedical model, which predominantly views the menstrual experience as a negative one [70]. However, if digital health and evolutionary ecologists were to meet to co-design research to understand the ultimate causes of variation in menstrual cycling, it might help the mHealth industry get closer to their goals of providing insights back to their users.

Can the concept of cyclical female health be extended to cyclically reproducing species? To count immunity as "cyclical", host defence must vary between reproductive and non-reproductive states. Such a trade-off between the functions of immunity and reproduction, which depends on the energetic costs of female reproduction, is well-documented in female insects [71] and birds [72], but less so in mammals [73,74] and reptiles [75]. Recent studies suggest that female health might be cyclical in wild baboons, where wound healing is found to be slower among lactating females as compared with pregnant or cycling females [74], and in the viviparous garter snake [75]. No such trade-off was observed in voles [73], however. Further studies on the interactions between immunity and female reproduction are warranted to evaluate the generality of the concept.

Shifting towards a cyclical model for female health (Box 4) will contribute new questions for biomedicine, public health and evolutionary ecology. During which phase of the menstrual cycle is it best to administer chemotherapy, vaccination and other medicine? Should the cycling life-history be considered when calculating chronic disease risk? What is the impact of changing patterns of menstrual cycling for female social behaviour, pathogen transmission and evolution? For decades, research has shied away from tackling female health because of its complexity. Using the new data collection tools available in humans and addressing the sex bias in animal studies [76], now is the time to face that complexity.

**Box 1: Evolutionary hypotheses for the evolution of menstruation**

Menstruation is rare in mammals (Figure I) and potentially costly (e.g. loss of tissue, narrowing of the fertility window). So why has menstruation evolved at all?

*Adaptive hypotheses*

Worthman proposed that menstruation evolved to signal fertility. This view was rejected because various menstruating primates exhibit overt menses despite the presence of obvious sexual swellings [77]. Profet proposed that menstruation is an adaptation to cleanse the uterus from pathogens brought about by sperm. This hypothesis predicts a link between promiscuity, the degree of uterine infection and blood volume. However, Strassman found that the copiousness and the timing of menstruation were not associated with the risk of infection among primates [77]. Finn also queried why non-mammalian species susceptible to pathogens through internal fertilization have not evolved menstruation [78]. Finally, Clarke proposed that menstruation evolved to rid the body of abnormal embryos [in ,79]. However, the idea that menstruation is adaptive is challenged by the argument that menstruation is rarely observed in natural fertility populations as most females are pregnant or exclusively breastfeeding. Yet, in high fertility human populations, menstrual rituals and taboos are nevertheless common cultural features [13].

*By-product hypotheses*

Strassman hypothesised that menstruation is a by-product of endometrial regression; endometrial regression would save energy compared to the maintenance of the state of differentiation required for implantation [7]. However, maintaining the endometrium in a differentiated state indefinitely would prevent the conditions necessaries future pregnancies [reviewed in ,4]. Finn proposed the alternative hypothesis that menstruation is a by-product of an implantation reaction evolved to protect the uterus from an invasive trophoblast. In menstruating species, the implantation reaction (i.e. spontaneous decidualization or SD) mobilizes the immune system and takes place even in the absence of fertilisation. That menstruation is a response to a stimulus to the endometrium in the absence of fertilisation has been experimentally tested in mice, where menstruation can be induced [in ,79]. However, empirical support for the link between SD and trophoblast invasiveness requires detailed phylogenetic analyses [4]. Most recently, Macklon and Brosens argued that SD evolved in the face of genetic abnormalities, which are thought to be more common in species exhibiting extended copulation due to ageing gametes [1,2]. In line with this, women with impaired SD exhibit both high fecundity and repeated pregnancy failure [10].

**Figure I: The phylogeny of menstruation in placental mammals.** Adapted from [4]. Menstruating species are observed in some species of primates, chiroptera and afrotheria (pink lineages). In primates, menstruation is observed in humans, apes and Old world monkeys, but is lacking in strepsirrhine (lemurs and lorises) and varies considerably within haplorhines [12]. The figure suggests that menstruation has been the subject of convergent evolution. Contemporary understanding is that menstruation has evolved as a by-product of uterine evolution in the face of invasive trophoblast (lacing in tarsiers for instance [12]) and genetic abnormalities [reviewed in ,4].

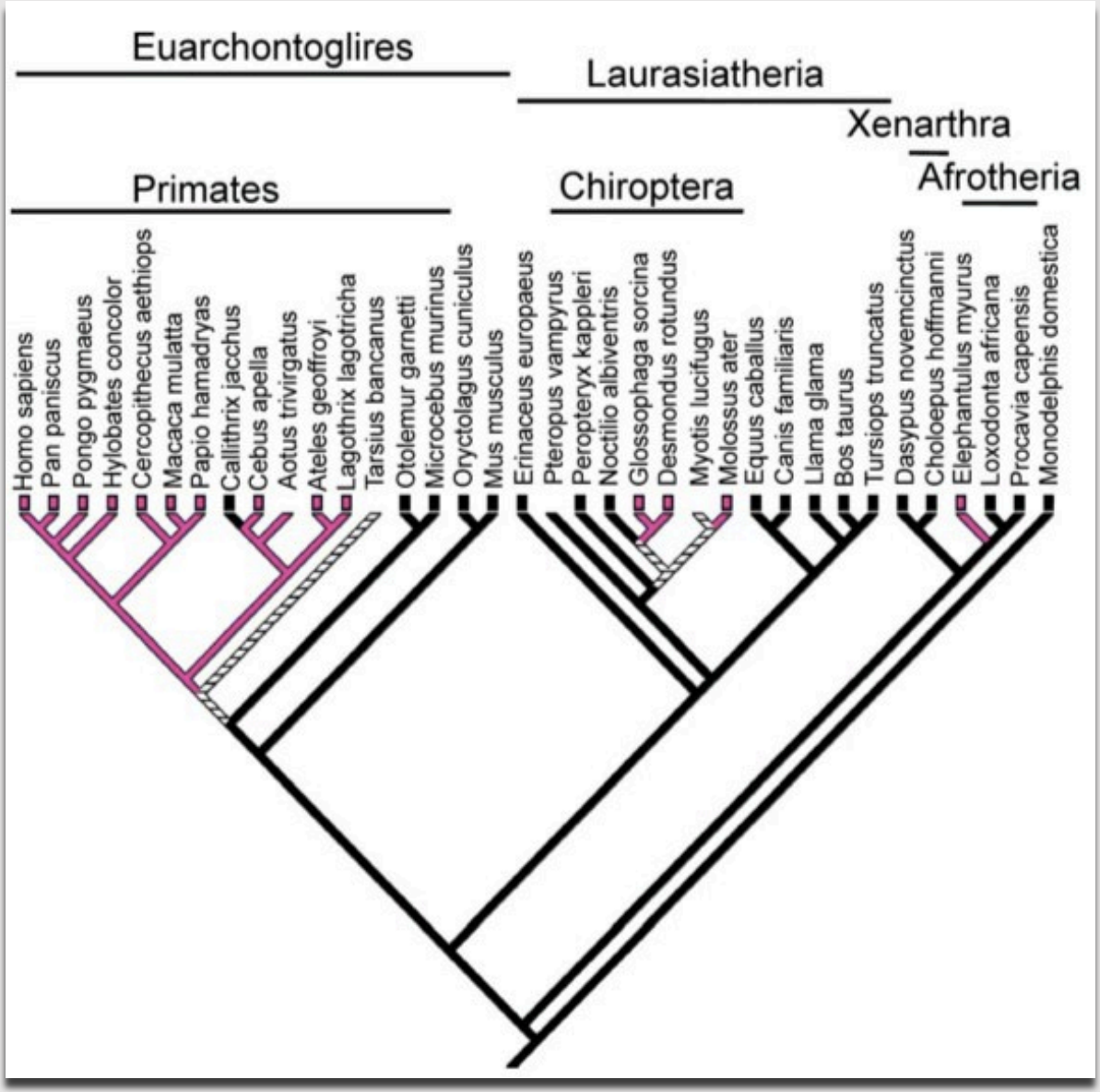

**Box 2: Hormonal contraception and health**

Hormonal contraception has been deemed the "innovation" of the 20[th] century for its role in women's empowerment and economic development. In addition to the benefits of controlling one's fertility, hormonal contraceptives enable women to reduce or suppress menstruation (amenorrhea), and the debilitating symptoms and conditions associated with it (e.g. PMS, migraine and pain, endometriosis) [80]. But what is the impact of hormonal contraception on disease susceptibility and severity?

*Hormonal contraception and reproductive cancers*

There is no evidence that the suppression of menstrual cycling has deleterious effects on non-smoking women. Rather research shows that prolonged use of hormonal contraception reduces the risk of reproductive cancers, in particular endometrial and ovarian cancers. Such beneficial outcomes are thought to result from a reduction of the number of mutations in epithelial cells following the suppression of cell division mechanisms otherwise activated for healing the ovary and the endometrium [81]. For explaining the ontogeny of ovarian cancer [17], the "incessant ovulation" hypothesis (i.e. the role of repetitive wounding of the ovaries) was initially favoured, but the "incessant menstruation" hypothesis (i.e. the role of repetitive exposure to iron-induced oxidative stress through retrograde menstruation) is gaining support. Given that menstruation is associated with biomarkers of systemic inflammation (Table1), it is possible that the "incessant menstruation" hypothesis could be extended to non-reproductive cancers.

*Hormonal contraception and infections*

Users of hormonal contraceptives are predicted to be at an increased risk of infections. Progesterone-based contraceptives like Depo-Provera (or DMPA, a commonly used contraceptive in sub-Saharan Africa) have routinely been utilized to facilitate infection in rodents [82] and studies on mice have shown that exposure to Depo-Provera is associated with poor response to herpes virus [83]. There seems to be an emerging consensus that progestin-based contraceptives increase the susceptibility to viral infections [84], and a recent meta-analysis of cross-sectional and longitudinal studies in humans suggests that DMPA adds to the risk of male-female HIV transmission [85]. It has also been found that users of hormonal contraception show a reduced immune response as compared to others. For instance, bacterial counts are higher in users of combined oral contraceptives compared to non-contraceptive users, after controlling for differences in sexual activity [19]. How contraceptive users and non-users differ in their response to vaccination is unknown.

**Box 3: The ecology of menstruation**

Understanding the ecological determinants of menstrual cycling diversity might shed new light on female immunity and behaviour, disease ecology and pathogen evolution. However, in both humans and non-humans, data on the ecology of menstruation and its impact on immunity are scant.

*Patterns of menstrual cycle diversity*

In non-human wild species, females spend most of their reproductive lives pregnant or lactating, and thus the event of menstruation is rare and data on menstrual cycle diversity are limited. In humans, according to a comprehensive review based on 21 studies across 11 populations, there is extensive variation between populations in cycle length (Figure II), period length (3.5 to 6 days), levels of steroids and probability of ovulation [86]. Such variation is partly determined by resource availability. For instance, progesterone levels respond to energy balance, and ovulation is suppressed in conditions of resource scarcity [13,30]. Poverty and rurality also influence variation in ovarian functioning, but no clear population patterns have emerged [86,87]. This gap in our understanding arises from methodological biases including too few cycles sampled, small sample sizes, differences in menstrual cycle phases definitions [86] and the underestimation of variation due to the exclusion of women with medically irregular cycles, about 30% worldwide [87]. Further, there are multiple factors that influence metabolic investment in reproductive functioning, including age, social capital, pathogen load, nutritional status and parity [13,30]. Digital health technology might offer a promising avenue given its potential to gather daily data over several cycles in various populations.

*Menstrual cyclicity in immune function*

Variation in the extent to which immunity changes over the menstrual cycle, a measure of reproductive investment, has been little researched to date. Recent studies in Western [27,31] and Bolivian contexts [22] indicate that cyclicity in inflammatory patterns only occurs in cycles that are ovulatory [22] and for sexually active women, suggesting that the menstrual modulation of immunity is a plastic response to the social ecology. Indeed, the degree of immune cyclicity is expected to reflect the optimal resource allocation decision between reproduction and immunity given individual circumstances (e.g. age, fat stores, social status and pathogen load). Future studies aiming at understanding the ecological determinants of diversity in menstrual immune cyclicity will shed new light on the evolution of female life-history in menstruating species.

**Figure II: Variation in the average cycle length across 11 human populations.** Reproduced from [87]. Existing data on global patterns of menstrual cycle diversity show that median cycle length can vary from 28 to 36 days, and that period length can vary from 3.5 to 6 days (not shown on graph, [86]). However, existing studies are too few, generally based on a small number of cycles and conducted in a limited number of populations. Reproductive ecologists have shown that on the one hand ovulation may be suppressed in conditions of resource scarcity, and on the other, that steroid hormone levels vary between populations without necessarily impairing the ovarian function [13,30]. The impact of poverty, rurality, social classes and infections on menstrual cycle characteristics (cycle length, period length, pre-menstrual symptoms, immune cyclicity) has yet to be fully uncovered.

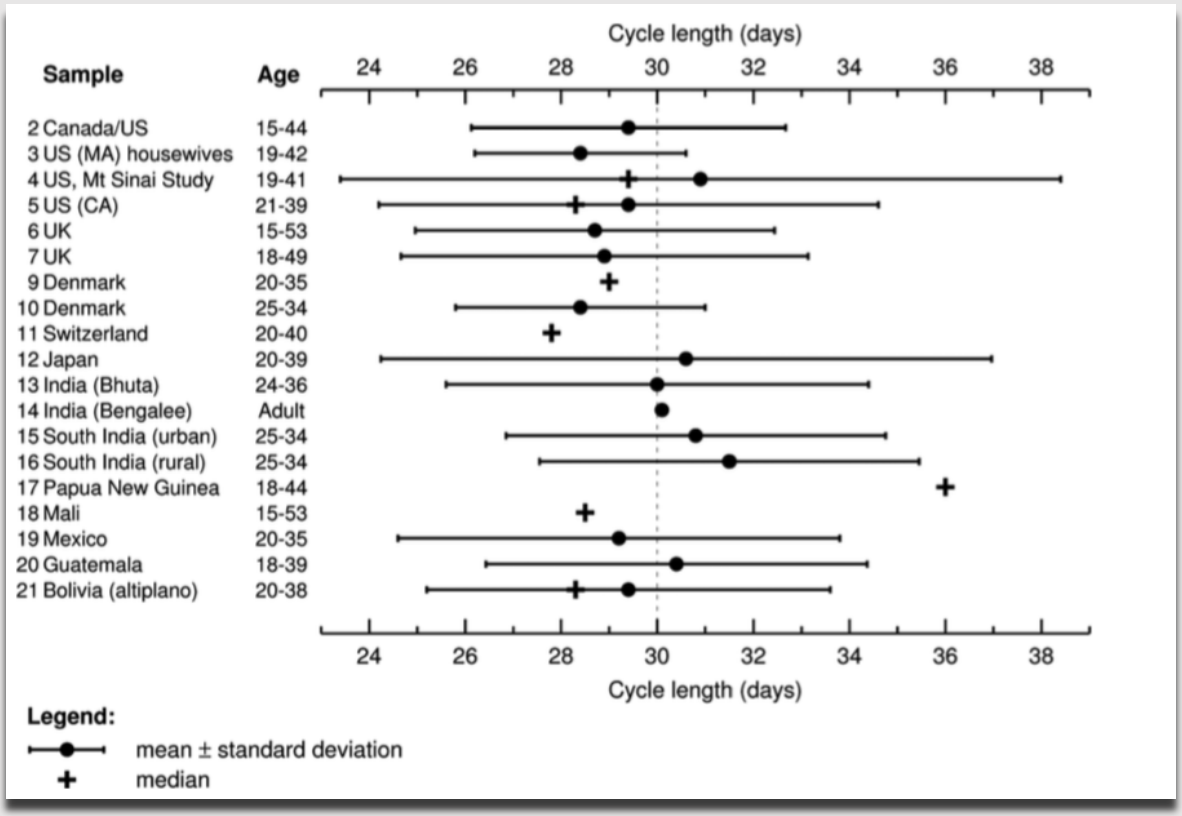

**Box 4: Towards an evolutionary cyclical model for women's health**

Most biomedical research on female health focuses on reproductive health, which, as pointed out by the social sciences, reduces female health to women's reproductive function [88]. Recent health policy reports thus call for a research program that moves beyond the study of reproduction to tackle the future challenges for women's health, cardiovascular diseases, cancers, autoimmune diseases, diabetes and mental health [89]. Arguably the global women's health agenda goes much beyond reproductive concerns, but we contend that it is mistaken to conceptualize women's health as separate from women's evolved reproductive system. From an evolutionary perspective, the regulation of the reproductive function is viewed as the very foundation of health. This is because reproductive features have a great impact on reproductive fitness and are thus particularly malleable to evolution by natural selection. This evolutionary force is constrained by resource allocation trade-offs [90] and thus energetic investment in reproductive function imposes constraints on the potential for the body to be "healthy", i.e. to invest in immuno-competence. The core insights to be taken from an evolutionary approach are that the body is best viewed as an evolved system of interconnected functions (growth, immunity, reproduction), and from the perspective of natural selection, health is only a means to the end of reproduction [90]. While it is well-known that the neuro-endocrine and immune systems are intimately connected [19], the biomedical sciences have yet to conceptualize reproductive health as foundational for non-reproductive health.

# Figure Legends

**Figure 1: The human menstrual cycle is characterized by both pro and anti-inflammatory processes.** The human menstrual cycle involves the tight regulation of inflammatory processes that enable the implantation of a healthy embryo. It is characterized by both an ovarian cycle (top graph) and a uterine cycle (bottom graph) that are regulated by hormones (middle graph). The figure represents a 28-day cycle idealized as an average cycle, however at least 30% of women experience shorter or longer cycles at some point in or during their entire reproductive history. By convention, the cycle starts with the first day of menstrual bleeding and ends the day before the onset of menstruation.

**The ovarian cycle (top graph).** At the beginning, the follicle stimulating hormone (FSH) is produced by the pituitary gland located at the base of the brain. It stimulates the development of follicules and oocyte maturation within the dominant follicule. The granulosa cells of the follicules start producing estradiol (E2), FSH levels decline, and gonadotropic cells in the anterior pituitary gland start producing the luteinizing hormone (LH). LH continues to rise as more E2 is being secreted; starting around day 10 of the menstrual cycle E2 sharply rises, which is followed by a massive LH surge, and a smaller spurt of FSH. The oocyte matures through completion of the first meiosis, commencement of the second meiosis and is arrested in metaphase, all of which happens within the 24 hours post LH surge. Next, the oocyte is physically expelled from the follicule, marking the ovulation (i.e. follicular rupture), which is followed by an acute inflammatory reaction to repair the damaged tissue. After ovulation, the empty follicule (corpus luteum) starts producing progesterone to enable the endometrium to prepare for implantation. At the later stage of the cycle, the corpus luteum produces estradiol as well.

**The uterine cycle (bottom graph)**. During the proliferative phase of the uterine cycle, estrogen promotes the repair and thickening of the inner lining of the uterus (endometrium); during the secretory phase, rising levels of progesterone (P) levels enable the decidual cell reaction (DCR), a process by which the endometrium prepares for implantation and which involves the recruitment of immune cells and the differentiation of endometrium cells into secretory cells producing proteins and growth factors. The DCR afford the uterus the capacity to (1) discriminate between healthy and compromised eggs due to an acute inflammatory reaction (ca. [Days 18-21]) and (2) allow for the implantation of a blastocyst due to a profound anti-inflammatory reaction (ca. days 23-27). Thus, the "receptive window" (days 20-24) is governed by both pro and anti-inflammatory processes [8]. If the oocyte is not fertilised, the corpus luteum degenerates, progesterone levels are not maintained, and an acute inflammatory reaction follows. The end result is menstruation, which corresponds to the third phase of the decidual cell reaction in a non-conceptive cycle [8].

**Figure 2: Menstrual cycling and non-reproductive health.** Data on non-human menstruating species are lacking thus only human data are reported. The faces represent the effect of the menstrual cycle phase on disease severity and/or susceptibility (happy: positive; sad: negative).

**(a) Menstrual cycle phases.** The menstrual cycle modulates disease susceptibility, development and symptoms. For instance, some symptoms are exacerbated during the inflammatory phases of the menstrual cycle (ovulation and menstruation), while susceptibility to infection is increased during the mid-luteal phase of the menstrual cycle, when high progesterone levels lead to an anti-inflammatory environment. Only two reviews have been conducted on the role of the menstrual cycle phase on various diseases [16,17] and many unknowns remain. For instance, given that estrogen both stimulates cell division and fluctuates during the menstrual cycle, does the menstrual cycle phase influence the division of cancer cells and how cancer progresses? Similarly, given high progesterone levels impair the response to infection, does the menstrual cycle phase influence the response to vaccination? **(b) Cycling life-history.** The cycling life-history, i.e., how many menstrual cycles women experience through their lives, is highly variable among women, depending on age at menarche, number of pregnancies, duration of breastfeeding, age at menopause and the use of hormonal contraceptives and hormonal replacement therapy. Given that menstruation is an acute inflammatory event [8], that progesterone is anti-inflammatory and that estrogen can be either inflammatory or anti-inflammatory depending on the dose, the cycling life-history is likely to influence the overall inflammatory load [55] of a woman and thus her risk of chronic and long-term illnesses. *E*: Estrogen; *P*: Progesterone.

**Table 1: Variation of an inflammatory biomarker (CRP) across the human menstrual cycle.** A dark box colour indicates a positive effect of the menstrual phase on CRP, light grey indicates a negative effect, and dashed patterns indicate non-significant relationships. *Pop.*: Population; *N*: Sample Size; *NC*: Nb of cycles per woman; *EFP*: Early Follicular Phase or Menses; *MFP*: Mid-Follicular Phase; *POP*: Peri-Ovulation Phase; *MLP*: Mid-Luteal Phase; *LLP*: Late-Luteal Phase; *OVL*: expected day of ovulation; *[ , ]* range of days during which sampling occurred; + /- days from the expected date of ovulation; *hs-CRP*: high-sensitivity CRP assays; *E*: Estrogens; *E2*: Estradiol; *P*: Progesterone; *DBS*: Dry Blood Spots; *na*: not available; *M,W,F*: Mondays, Wednesdays, Fridays.

| Pop. | N | NC | Age | Sampling CRP N samples/cycle | Ovulation detection | Days of sampling in each menstrual cycle phase | | | | | Comments | Ref. |
|---|---|---|---|---|---|---|---|---|---|---|---|---|
| | | | | | | EFP | MFP | POP | MLP | LLP | | |
| *PREDICTION for the menstrual modulation of CRP* | | | | | | Menses (HIGH) | 2ND WEEK (MEDIUM) | Ovulation (HIGH) | Selection 18-21 (HIGH) | Implantation 23-27 (LOW) | For healthy premenopausal individuals off hormonal contraceptives and with ovulatory cycles | |
| Polish | 25 | 1 | 20-40 | Urinary 7 | Mid-cycle E drop (saliva) | 3,4,5,6 | | | 21,22,23 | | Agricultural women; no differences between average CRP between phases; CRP negatively correlated with E2 and age at menarche | [54] |
| Austrian | 18 | 1 | 21-38 | Blood 3 | Ovulation test; LH surge | | 7 | +1 | | +9 | Median CRP levels increase by 44% at mid-cycle and by 31% in the LP | [91] |
| American | 8 | 1/2 | 21-47 | Blood (DBS) 12 | 2nd day of a 5 day drop in E/P | M,W,F | M,W,F | | M,W,F | | 10-fold increase in P -> 23% CRP increase; 10-fold increase in E -> 29% CRP decrease; menses->17% CRP increase | [92] |
| Swiss | 16 | 1 | 20-36 | Fasting Blood 14 | P levels on day 21 of preceding cycle | 3,5 | [8,16] | | 18,21,24 | 27 | 8 normal weight and 9 overweights; hs-CRP correlates negatively with E (beta =-0.23); hs-CRP highest at the beginning of the cycle | [93] |
| Swiss | 36 | 1 | 20-32 | Blood ca. 12 | LH detection test | Throughout (every other day, every day around ovulation) | | | | | CRP patterns show no significant changes during the cycle | [94] |
| Swedish | 102 | 1 | 31.7 (8.6) | Blood 2 | P > 15nM/L in the luteal phase; E higher in FP | [3,5] | | | | [22,25] | hs-CRP significantly higher in EFP as compared with MLP; 16 women with anovulatory cycles, did not change the results | [95] |
| Turkish | 27 | 1 | 25.9 (5.1) | Blood 2 | P > 3ng/mL in the luteal phase | [2,5] | | | | [21,24] | CRP higher in the EFP (1.8mg/L) as compared with MFP (0.7mg/L) | [96] |
| USA Mixed | 259 | 2 | 18-44 | Fasting Blood 8 | Ovulation test | na | na | na | na | | CRP highest during menses (median, 0.74 mg/L), decreased during the MFP, lowest on the expected day of ovulation (median, 0.45 mg/L), and increased in the LP. | [97] |
| Asian | 893 | 1 | na | Blood 5 | Date of last menstrual period + 13-16 | [1,6] | [7,12] | [13,16] | [17,22] | [23,31] | CRP higher in the EFP than in the MFP, POP, LLP. | [98] |
| Bolivian | 61 | 2 | 28.2 (3.95) | DBS 5-6 | Mean peak salivary P >110 pM/L; | | 6,7, [8-21], 22, 23, 24 | | | | Women living in demanding conditions; high early cycle CRP associated with anovulation; menstrual modulation of CRP depends on sexual activity | [22] |
| Italian | 18 | 1 | 21-44 | Serum 3 | P > 10ng/mL in days 19-23 | [2,3] | | [12,13] | | [23,24] | No menstrual variation of CRP among ovulatory cycles; higher CRP in the POP in ovulatory compared to anovulatory cycles | [99] |
| Swiss | 8 | 1 | 30.9 (3.3) | Serum 15 | P levels on day 21 of preceding cycle | [-15,-9] | [-8,-4] | [-3,-1] [+1,+3] | [+4,+8] | [+9,+14] | Hs-CRP independent of weight status and E & P serum levels Hs-CRP highest in EFP and associated with mood and behaviour | [100] |

# Glossary

**Adaptive immune response:** the second line of defences against pathogens (microorganisms harmful to the body). The adaptive immune response is highly specific, involves immunological memory and can provide long lasting protection.

**Biomarkers of inflammation:** measurable indicators that correlate with levels of inflammation. Common biomarkers include acute phase-proteins such as CRP and fibrinogen.

**Blastocyste:** the name given to the embryo between the 5th and 7th day after fertilization.

Compensatory prophylaxis hypothesis: it posits that menstrual variation in immunity triggers changes in disgust responses to pathogen cues in a way that compensates for the loss of immunity in the luteal phase of the cycle.

**C Reactive Protein (CRP):** an acute phase protein produced by the liver and adipocytes, and a marker of low-grade systemic inflammation and cardiovascular disease risk.

**Cycling life-history:** the lifetime number of cycles of menstruation. It is influenced by the age of menarche, the number of pregnancies, the duration of breastfeeding, the age at menopause and the duration of use of hormonal contraceptives and hormone replacement therapy.

**Cytokines:** small proteins involved in cell signalling and acting as immunomodulating agents. Interleukins are a sub-type of cytokines.

**Decidual cell reaction (DCR):** also referred to as the "implantation reaction". a reaction that enables the preparation of the endometrium for implantation of the embryo and a characteristic of menstruating species. After ovulation, rising levels of progesterone trigger cellular and vascular changes in the endometrium, which involve the recruitment of immune cells and the production of growth factors and proteins by endometrial cells. In the absence of successful conception, the decidual cell reaction ends with menstruation. to protect the uterus from an invasive trophoblast.

**Endometrium:** the inner lining of the uterus.

**Follicular phase:** a phase of the menstrual cycle. It starts on the first day of the menses and ends with ovulation.

**Immunocompetence:** the ability of the body to mount an immune response following exposure to a pathogen. The contrary of immunodeficiency.

**Inflammation:** a complex biological response to harmful stimuli and a part of the innate immune response. It is correlated with, but is not a synonym of, infection.

**Innate immune response:** the first line of defense against pathogens, which includes features such as the skin, the mucosa and immune cells. The innate immune response is activated immediately after an invading pathogen has been detected and acts within hours. This non-specific response can then activate a second line of defense: the adaptive immune response.

**Life history theory:** a branch of evolutionary theory which is concerned with the following question: for each sex, each life-stage and each ecology, how are resources allocated between the different fitness functions of growth, survival and reproduction in a way that optimizes lifetime reproductive success? Life-history theory predicts that when resources are limited, organisms will face resource allocation trade-offs between different fitness functions.

**Luteal phase:** a phase of the menstrual cycle. It occurs after ovulation and before the period start.

**Lymphocyte T helpers (Th):** immune cells which are part of the adaptive immune system. Lymphocyte T helpers help up or down regulate the immune response.

**mHealth:** Mobile health.

**Menstruation:** the cyclical shedding of the endometrium triggered by falling progesterone levels.

**Mitogenic:** encourages a cell to commence cell division

**Oocyte:** an egg cell, i.e. an immature ovum, produced by the ovary.

**Ovulation:** the release of an egg from one of the ovaries

**Parity:** the number of live births of a female.

**Phenotypic plasticity:** the ability of a genotype to express different phenotypes as a function of the environment.

**Pre-menstrual syndrome (PMS):** a chronic condition of unknown aetiology experienced by women before their menses. It is characterized by a wide variety of symptoms including depression, fatigue, cramps and headaches.

**Trophoblast:** tissue that forms the outer layer of the blastocyst and will later become a major part of the placenta.

**Virulence-transmission trade-off:** a hypothesis which posits that in the absence of vectors, pathogens face a trade-off between virulence and transmission [66]. This is because although virulence correlates with the fitness of pathogens, i.e. their ability to transmit between hosts, there is a point at which the cost of virulence for host death outweighs the transmission benefit.

**Waiting time to conception:** the interval between the resumption of cycling and the next conception.

Figure 1

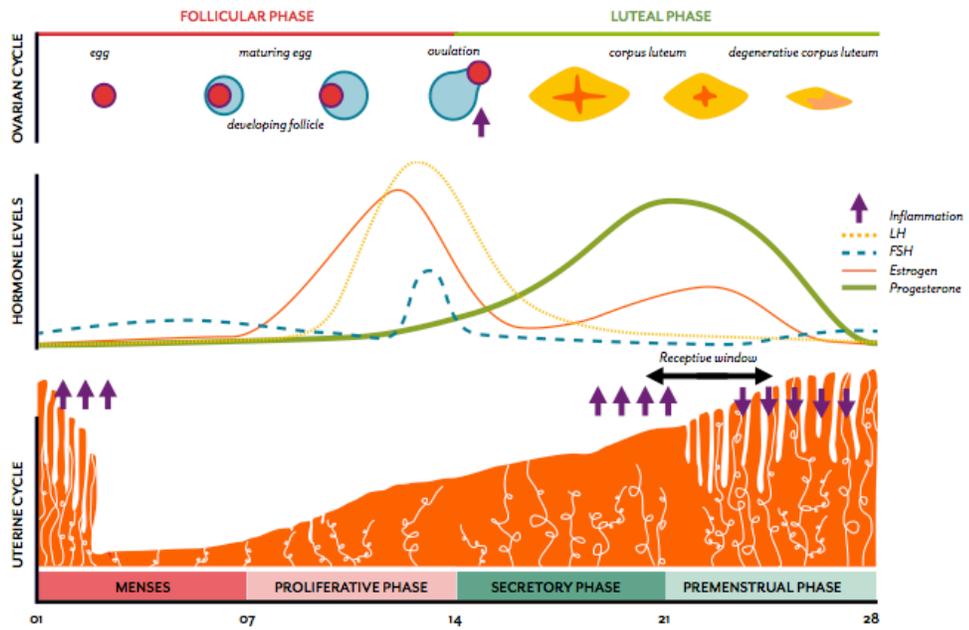

Figure 2

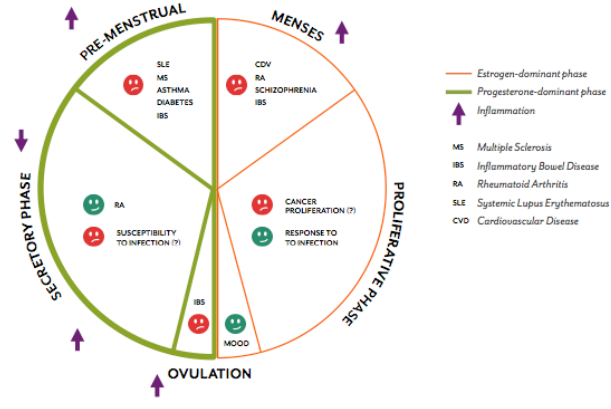

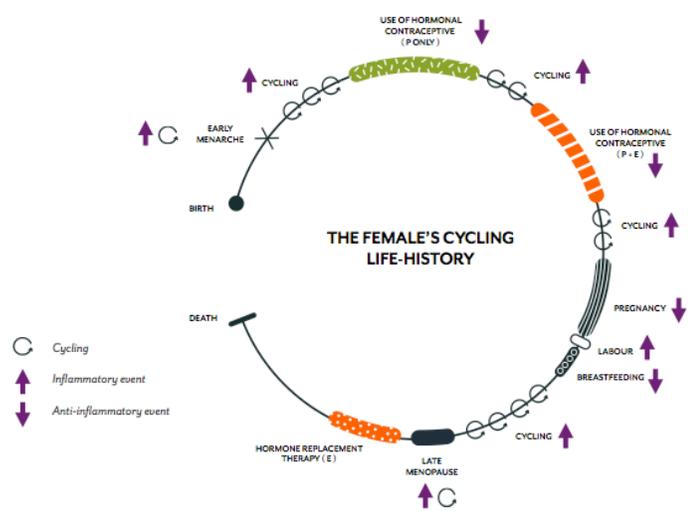